\documentclass[aps,prb,twocolumn,groupedaddress]{revtex4}
\usepackage{amsmath}
\usepackage{graphics}
\usepackage{graphicx}
\usepackage{epsfig}
\usepackage{amssymb}
\usepackage{amsmath}
\usepackage{amsfonts}

\begin{document}
\title{Quantum corrections to the phase diagram of heavy-fermion superconductors}
\author{Andr\'e S. \surname{Ferreira}}
\author{Mucio A. \surname{Continentino}}
\affiliation{Instituto de F\'{\i}sica, Universidade Federal Fluminense, \\
Campus da Praia Vermelha, \\
Niter\'oi, RJ, 24.210-340, Brazil.}
\email{andre@if.uff.br,mucio@if.uff.br}
\author{Eduardo C. \surname{Marino}}
\affiliation{Instituto de F\'\i sica, Universidade Federal do Rio
de  Janeiro,
\\
Rio de Janeiro, RJ, 21945-970, Brazil.}

\date{\today}

\begin{abstract}
The competition between magnetism and Kondo effect is the main
effect determining the phase diagram of heavy fermion systems. It
gives rise to a quantum critical point which governs the low
temperature properties of these materials. However, experimental
results made it clear that a fundamental ingredient is
missing in this description, namely superconductivity. In this
paper we make a step forward in the direction of incorporating
superconductivity and study the mutual effects of this phase
and antiferromagnetism in the phase diagram of heavy fermion
metals. Our approach is based on a Ginzburg-Landau theory
describing superconductivity and antiferromagnetism in a metal
with quantum corrections taken into account through an effective
potential. The proximity of an antiferromagnetic instability extends
the region of superconductivity in the phase diagram and
drives this transition into a first order one.
On the other hand superconducting quantum fluctuations near
a metallic antiferromagnetic quantum critical point gives rise to a first
order transition from a low moment to a high moment state in the antiferromagnet.
Antiferromagnetism and superconductivity may both collapse at a quantum bicritical point
whose properties we calculate.

\end{abstract}
\maketitle

\section{Introduction}

Competition between superconductivity (SC) and magnetism
plays an important role in determining the physical properties of strongly correlated
superconducting materials, such as high-T$_{c}$ superconductors~\cite{Nagaragan,Cho}
and heavy fermions~\cite{H,Thompson,Petrovic,Mathur,nature}. The study of the
phase diagrams of such materials suggests that these phases are intrinsically related.
This relation and the interface between these
phases are among the fundamental issues that have not yet been clarified in
condensed matter physics. In particular, the case of high-$T_{c}$ oxides represents a
formidable problem as the nature of the normal phase itself is
not  understood.

The study of heavy fermion materials has focused until recently on the
competition between long range magnetic ordering due to the RKKY interaction
and the Kondo effect. However, low temperature experiments in
these systems have shown that they can exhibit superconductivity near or in
coexistence with an antiferromagnetic (AF) phase close to a quantum critical point
(QCP)~\cite{steglich}. This can be conveniently tuned by varying some element concentration
or pressure~\cite{Mathur}. In distinction with the high-$T_{c}$ materials
in the case of heavy fermions we have a quite reasonable understanding
of their normal phase. Below the coherence line $T_{coh}$ this is
essentially a strongly correlated, nearly antiferromagnetic, Fermi liquid \cite{tcoh}.
Above this line and in particular at the QCP, the non-Fermi liquid
behavior is well understood in terms of quantum criticality
and the associated critical exponents~\cite{Mucio1}.

In this paper our aim is to investigate the mutual effects of
superconductivity and antiferromagnetism at zero temperature and
how this determines the phase diagram of three dimensional heavy
fermions (d=3). Our knowledge of the metallic phase in these
materials provides a solid starting point for the approach we use.
We consider a Ginzburg-Landau functional which contains the
magnetic and superconducting order parameters and their coupling.
We apply the effective potential method used in quantum field
theory~\cite{Jona,Coleman} to calculate the quantum corrections to
this classical action. This method is implemented up to one loop
order and requires the knowledge of the propagator solutions of
the quadratic functionals associated with the decoupled fields. In
such approach, the normal, metallic paramagnetic phase is
described by the usual dissipative propagator that takes into
account the dynamics of the spin fluctuations (paramagnons) in the
nearly antiferromagnetic metal~\cite{Hertz,millis}.

We investigate several phase diagrams in $d=3$. A possible scenario is that of a quantum
bicritical point separating a metallic antiferromagnet from a superconducting
phase whose universality class has been identified~\cite{andre}. The model
also describes the case where the AF and SC transitions occur at distinct quantum
critical points (QCPs), so that, there is a normal state between the AF and SC
phases. The proximity to an AF instability drives the SC transition to the
normal state into a first order one and enlarges the region of the phase
diagram where superconductivity exists~\cite{andre}.
On the other hand superconducting fluctuations drive an
AF-QCP into a discontinuous transition and give rise to a new AF phase with a reduced value of
the order parameter.
We present the Ginzburg-Landau model and the calculation of the effective
potential in
detail since, as far as we know, this method has not been sufficiently explored in condensed matter
physics. Although the model refers specifically to the coupling between AF and SC order parameters,
the results obtained
should be considered more general, independent of the specific nature of the instabilities, as long as,
the dynamics are similar. For example, a structural instability associated with a soft optical phonon mode
has a dynamic description which is similar to that given to the superconductor  here.

It is interesting that recent experiments in heavy fermions
suggest the existence of intrinsic inhomogeneities in these materials
near their magnetic quantum critical point~\cite{Flouquet}.
Our model provides a natural
explanation for these inhomogeneities, attributing them to the  weak first order transitions and
the associated phenomena of phase coexistence and spinodals arising from the
quantum corrections  due to the coupling between different order parameters.

We point out that similar models describing the competition between
superconductivity and antiferromagnetism have  been
proposed~\cite{Zhang,demler,Sigrist}. The emphasis however has been on the
classical aspects of this phenomenon and their aim was to describe
high-$T_{c}$ superconductors. The validity of this approach for this class of systems is now
subject of intense debate~\cite{Anderson,debate}.

\section{Paramagnon and superconducting propagators}

The model Ginzburg-Landau action  contains three real fields. Two
fields, $\phi_{1}$ and $\phi_{2}$, correspond to the two
components of the superconductor order parameter (the ground state
wave function, for example). The other field $\phi_{3}$, for
simplicity represents a one component antiferromagnetic order
parameter. The results are immediately generalized to a three
component AF vector field, with the unique consequence of changing
some numerical factors as discussed below. In order to include
quantum fluctuations by the effective potential method it is
convenient to describe the quadratic parts of the action by the
associated propagators. The propagator associated with the free,
quadratic action of the superconductor can be directly obtained
from a quantum Ginzburg-Landau action~\cite{Mucio2}. However,
finding the appropriate form of the superconductor propagator is a
difficult problem since the nature of the fluctuations driving the
superconducting transition can change considerably the quadratic
part of this generalized action. For metallic host where pair
breaking interactions due to magnetic impurities destroy
superconductivity the quadratic action has been obtained in
Ref.~\cite{Ramazashvili} and is associated with a dynamic exponent
$z=2$. The case of BCS superconductors in which any attractive
interaction $U$ makes the system superconducting at T=0 has been
studied in Ref.~\cite{Kirkpatrick}. In this case the form of the
action directly reflects the fact that the quantum critical point
occurs for $U=0$. Here we are interested in pure systems, so that
criticality is achieved by pressure and impurities are not
considered. Also, in our case, superconductivity is certainly
non-BCS and finding the correct propagator is in itself a very
difficult problem. In order to make progress, we choose to
consider the simplest generalization of the classical
Ginzburg-Landau action,

\begin{equation}
\label{propagator1}G_{0}(k)=G_{0}(\omega,\boldsymbol{q})=\frac{i}{k^{2}-m^{2}}
\end{equation}
where $k$ is a four-vector $(\omega,\boldsymbol{q})$ with
$k^{2}=\omega ^{2}-q^{2}$ and we took the Fermi wave vector $q_F=0$.
This propagator assumes the existence of a gap or pseudogap in the excitation spectrum
of the phase precursive to the superconducting one. There is plenty of evidence for such
pseudogaps in High-Tc
superconductors~\cite{pseudo} and also now in heavy-fermions systems as shown in recent
experiments~\cite{pseudohf}. Attractive Hubbard models also show normal
phases with charge ordering and a gap for excitations that
vanishes in the superconductor quantum critical point~\cite{Micnas}.

We work in Minkowsky space since the present superconductor propagator is
Lorentz invariant and we want to obtain the effective potential as in quantum field
theory, following closely the work of Coleman and Weinberg~\cite{Coleman}. It's clear
from Eq.~(\ref{propagator1}) that {\em time} acts as an extra direction on the same
footing as the spatial ones, so that, the associated dynamic exponent is $z=1$.

The quadratic functional associated with the magnetic part represented by the field
$\phi_{3}$,  the sub-lattice magnetization,
takes into account the dissipative nature of the
paramagnons near the magnetic phase transition. This approximation to the full
paramagnon propagator can be found in detail in the work by Hertz~\cite{Hertz}
and here we present a brief derivation. The expansion of the magnetic
free-energy functional in a power series of the order parameter can be written
as \begin{widetext}
\begin{eqnarray}
S_{mag}(\phi_3)&=&\int d^4k\hspace{.1cm} v_2(k)|\phi_3(k)|^2 +
\nonumber \\ && + \int d^4k_1d^4k_2d^4k_3d^4k_4\hspace{.1cm}
v_4(k_1,k_2,k_3,k_4)\phi_3(k_1)\phi_3(k_2)\phi_3(k_3)\phi_3(k_4)\delta(k_1+k_2+k_3+k_4)
+ \ldots \label{Smag}
\end{eqnarray}
\end{widetext}
and the coefficients $v_{m}$ are proportional to a loop of $m$
electron propagators~\cite{Hertz2}. For a Hubbard interaction $U$
the susceptibility $\chi$ calculated in RPA is
\begin{equation}
\mbox{Im}\chi(\omega,\boldsymbol{q})=\mbox{Im}\left( \frac{\chi_{0}
(\omega,\boldsymbol{q})}{1-U\chi_{0}(\omega,\boldsymbol{q})}\right)
\label{RPA}
\end{equation}
where $\chi_{0}$ is the susceptibility of the free-electron
model~\cite{Lindhard}. This susceptibility gives a good representation of the
paramagnon propagator associated with the quadratic coefficient $v_{2}$. If we
are near a magnetic instability we can expand $\chi_{0}$ for long wavelengths
and low frequencies. Substituting this expansion in the RPA form of the susceptibility
Eq.~(\ref{RPA}) and choosing the units appropriately we can
write the AF paramagnon propagator as
\begin{equation}
\label{propagator2}
D_{0}(\omega,\boldsymbol{q})=\frac{i}{i|\omega|\tau
-q^{2}-m_{p}^{2}}
\end{equation}
where $\tau$ is a characteristic relaxation time and $m_{p}^{2}$ is related to
local Coulomb repulsion  $U$ and the density of states at the Fermi
level $N(E_{F})$  by
\begin{equation}
m_{p}^{2}=1-UN(E_{F}).
\end{equation}

The quadratic parts of the effective functional are then
completely characterized by the propagators of
Eq.~(\ref{propagator1}) and Eq.~(\ref{propagator2}). Notice that the magnetic propagator,
Eq.~(\ref{propagator2}), is sufficiently general to be associated
with a Hamiltonian which preserves rotational
invariance~\cite{Hertz}. In this case $\phi_3$ is replaced by a
vector and in the second term of Eq.~(\ref{Smag}) the quartic
interaction is written in the form
$(|\boldsymbol{\phi}_3|^2|\boldsymbol{\phi}_3|^2)$ as appropriate
for vector fields. The use of a vector field modifies the
effective potential only by numerical factors as we show in the
next sections.

\section{Classical potential}

The classical part of the potential is given by,
\begin{align}
\label{pot}
V_{cl}(\phi_{1},\phi_{2},\phi_{3})=\frac{1}{2}m^{2}(\phi_{1}
^{2}+\phi_{2}^{2})+\frac{1}{2}m_{p}^{2}\phi_{3}^{2} +\nonumber\\
+ V_{s}(\phi_{1},\phi_{2}) + V_{p}(\phi_{3}) + V_{i}(\phi_{1},\phi_{2}
,\phi_{3})
\end{align}
where the self-interaction of the superconductor
field is,
\begin{equation}
V_{s}(\phi_{1},\phi_{2})=\frac{\lambda}{4!}(\phi_{1}^{2} + \phi_{2}^{2})^{2}
\end{equation}
and that of the antiferromagnet is given by,
\begin{equation}
V_{p}(\phi_{3})=\frac{g}{4!}\phi_{3}^{4}.
\end{equation}
The coupling $g$ above is related to the coefficient $v_{4}$ in the expansion of
Eq.~(\ref{Smag}) with all dependence on the four-vector $k$ ignored. Finally,
the last term is the (minimum) interaction between the relevant fields,
\begin{equation}
V_{i}(\phi_{1},\phi_{2},\phi_{3})=u(\phi_{1}^{2}+\phi_{2}^{2})\phi_{3}^{2}.
\end{equation}
This term is the first allowed by symmetry on a series expansion
of the interaction. It is possible that in regions of the phase
diagram where phase coexistence between superconductivity and
antiferromagnetism occurs higher order terms should be included in
the free energy. However, we will not consider the case of
coexistence in the present work. Notice that for $u>0$, which is
the case here, superconductivity and antiferromagnetism are in
competition and the possibility of AF acting as a pair breaking
perturbation is considered.

The minima of the classical potential
yield the classical ground states of the system~\cite{Sigrist}.
The different
possibilities for the phase
diagrams described
by the classical action at mean field level~\cite{Sigrist,Zhang,Mathur,pagliuso} are
shown schematically in
Fig.~\ref{possible diagrams}. The $T=0$ transitions are all continuous, second order phase transitions.
\begin{figure}[th]
\begin{center}
\includegraphics[height=2.3cm]{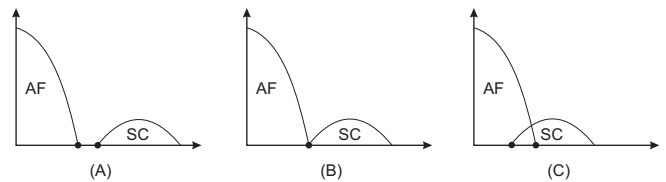}
\caption{Possible classical temperature phase diagrams for a heavy fermion system (schematic).
AF and SC refer to the superconductor and antiferromagnetic phases respectively. } \label{possible
diagrams}
\end{center}
\end{figure}
Notice that in Eq.~(\ref{pot}) there are two {\em mass} terms,
$m^{2}$ and $m_{p}^{2}$, which represent respectively, the
distance to the superconductor and the antiferromagnetic quantum
critical points. These two terms appear in the quadratic parts of
the action and are already included in the propagators
of~Eq.~(\ref{propagator1}) and~Eq.~(\ref{propagator2}) to be used
in the computation of the effective potential below.

The classical
potential can  be generalized to a rotational invariant form
replacing the scalar field $\phi_3$ by a vector. As discussed
before the self-interaction is then of the form
\begin{equation}
V_{p}(\phi_{3})=\frac{g}{4!}|\boldsymbol{\phi}_3|^2|\boldsymbol{\phi}_3|^2
 = \frac{g}{4!}\sum_{i,j}\phi_{3,i}^2\phi_{3,j}^2
\end{equation}
where $i$ and $j$ label the components of the vector. For this
generalized potential, however, the minima depend only on the
modulus of the vector $\boldsymbol{\phi}_3$ and the use of this
vector field brings only slight modifications to the effective
potential.

In the next
sections we obtain quantum corrections to the classical potential. These corrections can modify
significantly the physics of the problem and the phase diagrams changing, for example, the order of the
transitions.

\section{One loop effective potential}

\begin{figure}[tbh]
\begin{center}
\includegraphics[height=1.2cm]{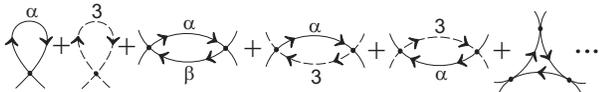}
\end{center}
\caption{One loop diagrams. The superconductor fields are represented by
$\alpha\mbox{ or } \beta=1,2$. The doted line represents the unusual
propagator of Eq.~(\ref{propagator2}).}
\label{fig loops}
\end{figure}

The first quantum correction to the potential can be obtained by
the summation of all one loop diagrams~(Fig.~\ref{fig loops}).

We apply the general method proposed by Coleman~\cite{ColemanBook}
with minimum modifications to account for the different nature of the propagators
in our problem. The sum over the
field indices can be easily done  if we define a vertex matrix $M$, given by
\begin{equation}
\label{M}[M]_{lm}=-iK^{l}_{0}\frac{\partial^{2}V_{cl}}{\partial\phi
_{l}\partial\phi_{m}}\Big|_{\{\phi\}=\{\phi_{c}\}}
\end{equation}
and then take the trace. In Eq.~(\ref{M}) the propagator (${K^{l}_{0}=G_{0}
\mbox{ or } D_{0}}$) of Eq.~(\ref{propagator1}) or Eq.~(\ref{propagator2}) is
incorporated in the definition of the matrix. We draw the loops with arrows
and choose the outgoing propagator of each vertex to be included in the associated
element. The matrix $M$ is then obtained deriving the classical potential with respect
to the fields $\{\phi\}$ and taking the values of these derivatives at the classical values of the fields,
$\{\phi_{ic}\}$. The sum of diagrams with the correct Wick factors is formally
done in momentum space and using the property of the trace
\begin{equation}
\mbox{Tr}[\ln(1-M)]=\ln\mbox{det}[(1-M)],
\end{equation}
we get
\begin{equation}
V^{(1)}[\phi_{c}]=\frac{i}{2}\hbar\int d^{4} k \ln\det\left[ 1-M(k)\right].
\end{equation}
The $3\times3$ matrix $M$ can be simplified if we choose the classical minimum of
the superconductor fields imposing ${\phi_{2c}=0}$ (this can be done because
the minimum depends only on the modulus $\phi_{1c}^{2}+\phi_{2c}^{2}$). Hence,
rotating to Euclidean space, so that, $k^{2}=\omega^{2}+q^{2}$ and using
$\hbar=1$ units the first quantum correction can be written as
\begin{widetext}
\begin{eqnarray}
\label{oneloop}
V^{(1)}(\phi_{1c},\phi_{3c})&=&\frac{1}{2}\int\frac{d^4k}{(2\pi)^4}\left\{
\ln\left(1+\frac{A(\phi_{1c},\phi_{3c})}{k^2+m^2}\right)+
\ln\left[\left(1+\frac{B(\phi_{1c},\phi_{3c})}{k^2+m^2}\right)
\left(1+\frac{C(\phi_{1c},\phi_{3c})}{|\omega|\tau+q^2+m_p^2}\right)+
\right.\right.\nonumber \\ \label{QC}
&& \left.\left.-\left(\frac{D^2(\phi_{1c},\phi_{3c})}{(k^2+m^2)(|\omega|\tau+q^2+m_p^2)}\right)
\right]\right\}
\end{eqnarray}
\end{widetext}
where
\begin{align}
A(\phi_{1c},\phi_{3c})=(\lambda/6)\phi_{1c}^{2}+2u\phi_{3c}^{2}\\
B(\phi_{1c},\phi_{3c})=(\lambda/2)\phi_{1c}^{2}+2u\phi_{3c}^{2}\\
C(\phi_{1c},\phi_{3c})=2u\phi_{1c}^{2}+(g/2)\phi_{3c}^{2}\\
D(\phi_{1c},\phi_{3c})=4u\phi_{1c}\phi_{3c}\label{D coef.}%
\end{align}
The total effective potential with first order quantum corrections is then given
by
\begin{equation}
V_{ef}(\phi_{1c},\phi_{3c})=V_{cl}(\phi_{1c},\phi_{3c})+V^{(1)}(\phi_{1c}
,\phi_{3c})
\end{equation}
where $V_{cl}$ is the classical potential of Eq.~(\ref{pot}) and
$V^{(1)}$ is the first quantum correction of order $\hbar$ of
Eq.~(\ref{QC}). In the subsequent sections we consider the effects
of these quantum correction  for the classical results. We give
special attention to the possibility of fluctuation induced
symmetry breaking and first order phase transitions at $T=0$. We
point out again that the results for an isotropic spin model are
similar since, when dealing with a vector magnetic order parameter
$\boldsymbol{\phi}_3$, we can use the fact that the classical
minima depend only on the modulus of this vector. The
modifications which arise in the final effective potential are
simply different numerical factors coming from new but equivalent
terms in Eq.~(\ref{oneloop}).

\section{Possible phase diagrams}

The ground state phase diagram of a heavy-fermion with AF and SC
phases can be obtained from the classical action by varying the
masses $m$ and $m_{p}$. The normal paramagnetic state has
$m_{p}^{2}>0$ and $m^{2}>0$ and for these values of the masses the
configuration $\phi_{1c}=\phi_{3c}=0$ minimizes the action. The AF
and SC states have $m_{p}^{2}<0$ ($\phi_{3c}\ne0$) and $m^{2}<0$
($\phi_{1c}\ne0$), respectively. The $T=0$ transitions between the
different phases,  which are tuned varying the masses, are up to
the classical level all continuous, second order transitions. The
different phase diagrams,  at mean field
level~\cite{Zhang,Mathur,pagliuso} were shown already in Fig.~\ref{possible
diagrams}. However, renormalization group studies of the classical
action yield that a bicritical point, as that of
Fig.~\ref{possible diagrams}B but occurring at finite
temperatures, is stable only if the number of components of the
superconductor ($n_S$) and antiferromagnet ($n_{AF}$) order
parameters~\cite{nelson} are such that $n= n_S + n_{AF} \le 4$.
For $n>4$ the SO($n$) bicritical point at $T \ne 0$ becomes
unstable and a phase diagram like that of Fig.~\ref{possible
diagrams}C may occur. The above constraint implies, for example,
that in a heavy fermion antiferromagnet of the Ising type, long
range magnetic order can not coexist with superconductivity ($\phi_1 \ne 0$ and $\phi_3 \ne 0$).
Notice that quantum corrections may destroy the classical SO($n$)
symmetry associated with the quantum bicritical point in
Fig.~\ref{possible diagrams}B due to the different dynamics of the
magnetic and superconductor fluctuations.

In the next sections, we study the quantum effects in two of the possible
phase diagrams of Fig.~\ref{possible diagrams}, namely, cases
(A) and (B). We begin considering case~(A) where
there is a normal phase separating the SC and AF phases and the transitions
occur at different QCPs. The quantum effects become important in the normal
region near both SC and AF phases where symmetry breaking and fluctuation induced
quantum first order transitions can occur. We distinguish here between the  case
the system becomes antiferromagnetic in the presence of superconducting fluctuations
from that where it becomes superconductor in the presence of antiferromagnetic fluctuations.

Case~(C) of Fig.~\ref{possible diagrams}, with some additional assumptions, can also
be described in our approach. In this case $\phi_3$ should be identified as the longitudinal
component of the sub-lattice magnetization with a relaxational dynamics described by the
propagator, Eq.~(\ref{propagator2}). Also we have to neglect spin wave excitations which,
at $T=0$, could be imagined as quenched
if, for example, there is a sufficiently large anisotropy gap in their spectrum.
In any case this situation has revealed up to now analytically
intractable. Although it is possible to perform an expansion of the last
logarithmic term in Eq.~(\ref{oneloop}) in powers of $u$, this
would be incompatible with the one-loop expansion which sums all
powers of this quantity when performing the relevant integrals
analytically, as we do below.

\section{Quantum effects in the normal phase between the AF and SC-QCPs}

\subsection{Fluctuation effects in the normal phase near superconductivity}

We are interested here in the superconducting-normal quantum phase transition and
therefore, near the SC state, we look for a partially symmetry broken phase with
$\phi_{1c}\neq0$ but $\phi_{3c}=0$. In this case $D(\phi_{1c},\phi_{3c})=0$
and the quantum correction given by Eq.~(\ref{QC}) can be written as
\begin{align}
V^{(1)}(\phi_{1c})=\frac{1}{2}\int\frac{d^{4}k}{(2\pi)^{4}} \ln\left(
1+\frac{(\lambda/6)\phi_{1c}^{2}}{k^{2}+m^{2}}\right) +\nonumber\\
+ \frac{1}{2}\int\frac{d^{4}k}{(2\pi)^{4}}\ln\left( 1+ \frac{(\lambda
/2)\phi_{1c}^{2}}{k^{2}+m^{2}}\right)  +\nonumber\\
+\frac{1}{2}\int\frac{d^{4}k}{(2\pi)^{4}}\ln\left( 1+\frac{2u\phi_{1c}^{2}
}{|\omega|\tau+q^{2}+m_{p}^{2}}\right) \label{pot1}
\end{align}
The first two integrations depend only on the modulus of the
four-dimensional vector $k$ and time enters as an extra dimension
as we are dealing with a Lorentz invariant case. This arises since
the QCP of the superconductor transition (SQCP) has an associated
dynamic exponent $z=1$. Therefore a cut-off regularization can be
done as usual~\cite{Coleman}. However, in the last integration we
have anisotropy between time and space since the dynamic exponent
which characterizes the scaling of time takes the value $z=2$.
Hence, if we use a cut-off $\Lambda$ to the momentum, the
correspondent frequency cut-off~\cite{Hertz} must be
$\Lambda^{z}=\Lambda^{2}$. The integrations in~Eq.~(\ref{pot1})
can be easily performed and the results expanded in powers of the
parameter $m^{2}$ supposing proximity to the superconductor
transition ($m^{2}\approx0$). The renormalization is done
beginning from the massless case $m^{2}=0$ and generalized for
small $m^{2}$ following closely the renormalization procedure for
the charged superfluid~\cite{Mucio2}. The effective potential
\begin{equation}
V_{ef}(\phi_{1c},\phi_{3c}=0)=V_{cl}(\phi_{1c},\phi_{3c}=0)+V^{(1)}(\phi_{1c})
\end{equation}
where $V_{cl}$ is the classical potential given by~Eq.~(\ref{pot}), is
\begin{widetext}
\begin{eqnarray} \label{pot2}
V_{ef}(\phi_c) \approx
\frac{1}{2}m^2\phi_{1c}^2+\frac{\lambda}{4!}\phi_{1c}^4+
\frac{\pi^2}{(2\pi)^4}\bigg[\frac{8}{15}(2u)^{5/2}\phi_{1c}^5
-\frac{8}{3}(2u)^{5/2}\langle\phi\rangle\phi_{1c}^4 +
\frac{4}{3}(2u)^{3/2}m_p^2\phi_{1c}^3-\frac{8}{3}m_p^3u\phi_{1c}^2\bigg].
\end{eqnarray}
\end{widetext}
Terms proportional to $\lambda^{2}$ and $m^{2}\lambda$ were neglected since
they must be much smaller than the classical term proportional to $\lambda$ in
the small coupling limit. The effective potential,~Eq.~(\ref{pot2}), contains
only the lowest order term in
$m_{p}^{2}$ since we are near both transitions and
consequently $m_{p}$ is also
small. The term between brackets is the first quantum correction of order
$\hbar$. As usual, the effective potential can be written as a function of its
extremum, $\langle\phi\rangle$, which determines if the system is in the normal
or in the broken symmetry superconducting phase. Higher powers in $u$ are not
neglected since we have no classical term proportional to $u$. Consequently,
even if $m^{2}>0$, balancing the classical $\lambda$ term with the $u$ terms
of the quantum correction, it is possible to obtain asymmetric minima for
$\phi_{1c}$, i.e., a symmetry breaking in the normal state induced by the coupling $u$. For small
masses $m_{p}$, independently of the value of $u$, the asymmetric minima are
close to the origin $\phi_{1c}=0$, as we can show by numerical inspection of
the full effective potential (this also justifies the expansion for small
$m_{p}^{2}$). Furthermore, with all minima close to the origin, we can neglect
$\phi^{5}$ terms that cause instabilities to the potential (fortunately these
terms become relevant far from the origin and hence away from the region of
validity of the loop expansion).

For convenience, in Eq.~(\ref{pot2}) we have introduced the
separation $\Delta^{2}= m^{2} + m_{p}^{2} >0$ (see
Fig.~\ref{phasediagram}) between the magnetic and superconductor
$T=0$ transitions and study the phase diagram as $\Delta^{2}$ is
reduced. Notice that $\Delta^{2}$ measures the distance between
the second order mean field QCPs in Fig.~\ref{possible diagrams}
and is also taken as a small quantity besides $m^{2}$ and
$m_{p}^{2}$. Now, studying the minima of the effective potential
we see that the quantum corrections induce symmetry breaking and
the region where superconductivity is found in the phase diagram
is extended. The shift of the superconductor QCP occurs towards
the antiferromagnetic quantum critical point (AF-QCP), but in the
paramagnetic phase. The analysis of the extrema of the potential
can be easily carried out and has been presented
before~\cite{andre}. We also find that the quantum fluctuations
changes the nature of the superconducting transition to a first
order one.  The {\em latent heat} and spinodal points for this
transition have also been calculated~\cite{andre}. Between the two
spinodals there is an interval of coexistence of the
superconducting phase with regions of strong antiferromagnetic
fluctuations due to the occurrence of metastable minima.

\subsection{Fluctuation effects in the normal phase near the AF-QCP}

\begin{figure}[thth]
\begin{center}
\includegraphics[height=3.9cm]{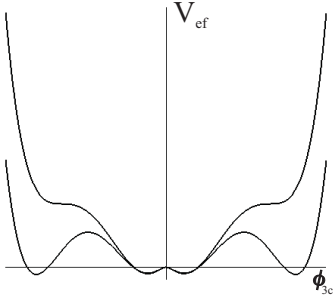}
\end{center}
\caption{ New minima appear in the potential for $um^2 \ne 0$. The
effective potential is shown here for two situations: for
$M_p^2=M_p^{c2}$, where the first order transition from LMAF to
SMAF occurs and these two states become degenerate and for the
spinodal point at which the LMAF becomes unstable inside the SMAF
phase. } \label{new transition}
\end{figure}

\begin{figure}[thth]
\begin{center}
\includegraphics[height=6.5cm]{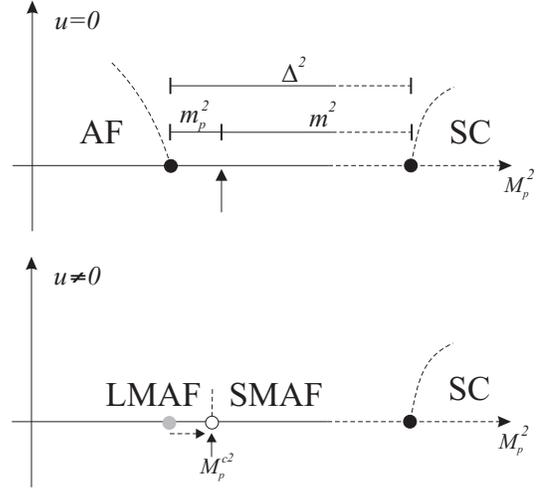}
\end{center}
\caption{ Phase diagrams for $u=0$ and $u \ne 0$. $\Delta^{2}$
is the distance between the AF and SC QCPs. $m_p^2$ and $m^2$ are
the distances from the point the system is actually probed to the
AF and SC QCPs respectively. When $M_{p}^{2} \approx m_p^2$ is reduced, i.e.,
the system moves away from the SC-QCP, there is a first order
transition at $M_{p}^{c2}$ between two AF phases with different
values of the
order parameter.}
\label{phasediagram}
\end{figure}

We follow the same procedure here as in the last section. Now we are
interested in the AF quantum phase transition and therefore we look for a partially
symmetry broken phase with $\phi_{3c}\neq0$ but $\phi_{1c}=0$. We have once
again $D(\phi_{1c},\phi_{3c})=0$ in the quantum correction given by
Eq.~(\ref{QC}) and the three resulting integrals have the same functional form
of Eq.~(\ref{pot1}) but with different $\phi_{c}$ dependence. Integration and
renormalization can be done as before beginning from the $m^{2}=0$ case. The
full effective potential
\begin{equation}
V_{ef}(\phi_{1c}=0,\phi_{3c})=V_{cl}(\phi_{1c}=0,\phi_{3c})+V^{(1)}(\phi_{3c})
\end{equation}
is similar to that obtained in the previous section except for
the renormalization counterterms. Terms
proportional to $u^{2}$ or $\lambda^{2}$ in the quantum corrections can not be neglected
since there are no $\lambda$ or $u$ terms in
the classical potential ($\phi_{1c}=0$). On
the other hand, there is a classical coupling $g$ and higher order terms in $g$
can be discarded. Notice that for $m^{2}=0$ the result is the same as for the charged
superfluid~\cite{Mucio2,Coleman},
\begin{align}
V_{ef}(\psi_{c},m  & =0)\approx\frac{1}{2}\left(  m_{p}^{2}-\frac{m_{p}^{3}
g}{12\pi^{2}}\right)  \phi_{3c}^{2}+\frac{g}{4!}\phi_{3c}^{4}+\nonumber\\
& +\frac{u^{2}}{8\pi^{2}}\phi_{3c}^{4}\left[  \ln\left(
\frac{\phi_{3c}^{2} }{\langle\phi_{3}\rangle^{2}}\right)
-\frac{25}{6}\right] \label{m=0 case}
\end{align}
and therefore there is a fluctuation induced quantum first order transition
varying $m_{p}^{2}$ for $g\sim u^{2}$. However,  the correct results in the present case
must be obtained for $m^{2}\neq0$ where superconducting fluctuations
are important but not critical. The effective potential to lowest order in powers
of $m^{2}$, which is also small but finite, is given by,
\begin{align}
V_{ef}  & \approx\frac{1}{2}M_{p}^{2}\phi_{3c}^{2}+\frac{g}{4!}\phi_{3c}
^{4}+\frac{u^{2}}{8\pi^{2}}\phi_{3c}^{4}\left[  \ln\left(  \frac{\phi_{3c}
^{2}}{\langle\phi_{3}\rangle^{2}}\right)  -\frac{25}{6}\right]  +\nonumber\\
& +\frac{um^{2}}{16\pi^{2}}\left[  \phi_{3c}^{2}+2\phi_{3c}^{2}\ln{\left(
\frac{2u\phi_{3c}^{2}}{\Lambda^{2}}\right)  }\right]  +\mathcal{O}
(m^{4})\label{m^2 expan.}
\end{align}
where $M_{p}^{2}$ is a renormalized magnetic mass parameter
\begin{equation}
M_{p}^{2}=m_{p}^{2}-\frac{m_{p}^{3}g}{12\pi^{2}}.
\end{equation}
We have now in Eq.~(\ref{m^2 expan.}) a new term proportional to $um^{2}$ and
since it depends on the cut-off $\Lambda$ another counterterm is needed. With a minimum
counterterm we get,
\begin{align}
V_{ef}(\phi_{3c})  & \approx\frac{1}{2}\left(  M_{p}^{2}+\frac{9u^{2}
\langle\phi_{3}\rangle^{2}}{2\pi^{2}}-\frac{1}{2}g\langle\phi_{3c}\rangle
^{2}\right)  \phi_{3c}^{2}+\nonumber\label{m^2<>0 case}\\
& +\frac{g}{4!}\phi_{3c}^{4}+\frac{u^{2}}{8\pi^{2}}\phi_{3c}^{4}\left[
\ln\left(  \frac{\phi_{3c}^{2}}{\langle\phi_{3}\rangle^{2}}\right)  -\frac
{25}{6}\right]  +\nonumber\\
& +\frac{um^{2}}{8\pi^{2}}\phi_{3c}^{2}\left[  \ln{\left(  \frac{\phi_{3c}
^{2}}{\langle\phi_{3}\rangle^{2}}\right)  }-3\right]
\end{align}
where other small quadratic terms in the field have been included
in the renormalized mass $M_{p}^{2}$. The coupling to the massive
superconductivity fluctuations changes dramatically the behavior
of the potential since now its second derivative at $\phi_{3c}=0$
is always negative, i.e., the origin is always a maximum for any
$um^{2}\neq0$. This coupling also gives rise to new minima near
$\phi_{3c}=0$ which move away from the origin as its strength is
increased~(Fig.~\ref{new transition}). It is interesting that as
$M_{p}^{2}$ is reduced, i.e., as the system moves away from the
SC-QCP towards the AF-QCP, it goes, through a quantum first order
transition at $M_{p}^{c2}$,  from a phase with a reduced value of
the antiferromagnetic order parameter to another with a larger
value of $\phi_{3c}$~(Figs.~\ref{new transition} and
\ref{phasediagram}).  The AF phase closest to the superconductor is always the
small moment one.

\begin{figure}[thth]
\begin{center}
\includegraphics[height=7.8cm]{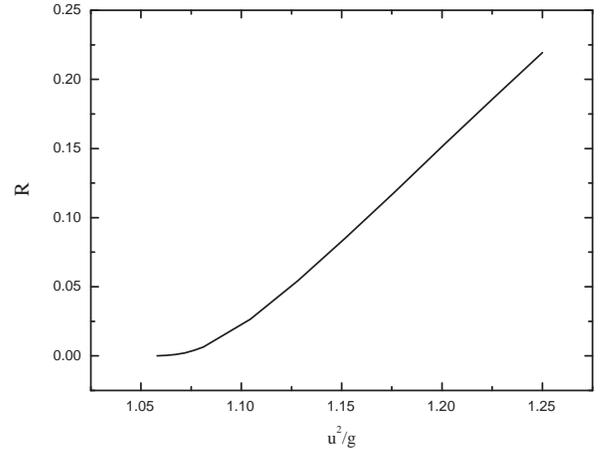}
\end{center}
\caption{The ratio $R=\langle \phi_c^{SM} \rangle/\langle
\phi_c^{LM} \rangle$ between the ground state sub-lattice
magnetizations in the phases SMAF and LMAF as a function of
$g^2/u$.} \label{ratio}
\end{figure}

In Fig.~\ref{ratio} the ratio $R$ of the
sub-lattice magnetizations in the small moment
antiferromagnetic~(SMAF) and large moment antiferromagnetic~(LMAF)
phases is plotted as a function of the relevant parameters, namely
the ratio $u^2/g$. Small moment antiferromagnetism (SMAF) is a
common feature in heavy fermion materials~\cite{buyers}. The most
common values for the ratio $R$ observed
experimentally~\cite{buyers,meguid} are of order $10^{-2}$,
$10^{-3}$ and correspond to values of $u^2/g \approx 1$. We have
found that the ratio $R$ is independent of the values of $g$ and
$\Delta$ being a function exclusively of the scaled variable
$u^2/g$.

There is an interesting similarity between the phase diagram of
the heavy fermion $YbRh_2Si_2$, for which a first order phase
transition from a low moment AF phase to a large moment magnetic
phase has been observed with increasing pressure \cite{meguid} and
the results obtained above. In $Yb$ based heavy fermions, pressure
(P) acts on the opposite direction it does on $Ce$ compounds
decreasing the ratio $J/W$ between the Kondo lattice parameters
\cite{Mucio1}. $YbRh_2Si_2$ at $P=0$ is a SMAF system with
$T_N=70mk$. Pressure increases $T_N$ and at $P_c \approx 10 GPa$
there is a first order transition to  a high moment state with
$\mu_{Yb} \approx 1.9 \mu_B$. On the other hand, negative
pressure, i. e., expansion of the lattice drives this system to an
antiferromagnetic quantum critical point \cite{trova}. It is an
exciting possibility that further expansion of the lattice would
give rise to superconductivity.

When  the separation~$\Delta^2$ between the classical $T=0$
critical points is reduced the magnetic moment in the SMAF goes
continuously to zero. In practice, when the separation is small
enough, the moment in the SMAF is so low that it can be identified
with the normal state. Then for $\Delta^2$ sufficiently small, but
still large enough to avoid the possibility of a direct transition
from AF to SC, there is a first order AF-normal transition. The
spinodal points and the energy equivalent to the latent heat can
be calculated exactly as in the charged
superfluid~\cite{Mucio2,Mucio3} since the term proportional to
$um^{2}$ can be neglected. Particularly, the effective potential
at the asymmetric minimum close to the first order transition can
be written as
\begin{equation}
V_{eff}(\langle\phi_{3c}\rangle)\approx\frac{1}{4}M_{p}^{2}\langle\phi
_{3}\rangle^{2}\left[  1-\frac{m_{c}^{2}}{M_{p}^{2}}\right]
\end{equation}
where the critical mass $m_{c}^{2}$ is
\begin{equation}
m_{c}^{2}=\frac{3u^{2}}{12\pi^{2}}\langle\phi_{3}\rangle^{2}.
\end{equation}
Therefore, the latent heat is given by the simple expression
\begin{equation}
L_{h}=\frac{1}{4}m_{c}^{2}\langle\phi_{3}\rangle^{2}.
\end{equation}
A similar phenomenon occurs for the first order SMAF-LMAF
transition. In this case there is also an associated latent heat
and the spinodal point corresponding to the limit of stability of
the large moment phase into the SMAF phase is shown in
Fig.~\ref{new transition}.

\section{Quantum bicritical point}

In this section we study the possibility of a quantum bicritical
point (QBP) as shown in Fig~\ref{possible diagrams}B. The quantum
corrections can be calculated in the paramagnetic side and involve
terms of higher order in the couplings $u$, $\lambda$ and $g$. In
the limit of small coupling considered in this work, these
corrections are then of little importance, especially for the
quantum transition, since the classical part becomes much larger.
However, mass renormalization can change the properties of the
superconductor itself as, for example, it affects the constant
$\kappa$ which determines whether the system is a type I or type
II~\cite{andre}.

\begin{figure}[th]
\begin{center}
\includegraphics[height=4.5cm]{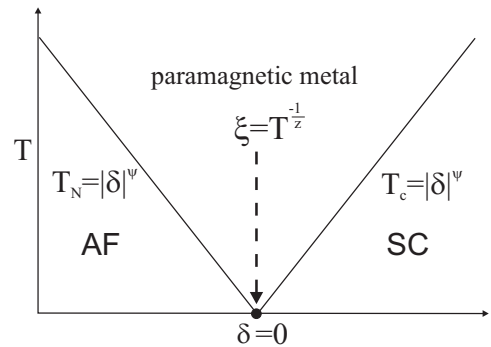}
\end{center}
\caption{A quantum bicritical point separating an antiferromagnet from a
superconductor. Both lines of finite temperature phase transitions rise with
the same exponent $\psi$. For d=3, $\psi=\nu z =1$.}
\label{bicri}
\end{figure}

The transition where both phases collapse is continuous, as
obtained classically and the analysis of this double critical
point can be done more easily using a quantum scaling
theory~\cite{Mucio4}. Classically this point has SO($n$)
symmetry~\cite{Zhang}, but when taking into account its quantum
character, as it occurs at $T=0$, there are different dynamics
related to the distinct phases on both sides of the transition. It
is easy to convince oneself that in the low frequency, long
wavelength limit, it is the slow relaxation dynamics of the
magnetic component with the associated $z=2$ dynamic exponent that
is relevant for the critical behavior. For $d=3$, assuming that a
unique correlation length diverges as $T$ is reduced at the QBP,
we obtain that both the antiferromagnetic ($T_{N}$) and
superconducting ($T_{S}$) critical lines of finite temperature
phase transitions rise with the distance $\delta$ to the QBP, as
$T_{N,S} \propto|\delta|^{\psi}$, i.e., with the same shift
exponent $\psi$. Furthermore, we find $\psi=\nu z$, where $\nu$
and $z$ are the correlation length and the dynamic exponents
associated with the QBP. Since the effective dimension
$d_{eff}=d+z=5$, we have $\nu=1/2$, such that the quantum
bicritical crossover exponent~\cite{Mucio4} $\nu z=1$.
Alternatively, the critical line can be written as
$\delta(T)=\delta(T=0)+uT^{1/\psi}$. For the correlation length
along the critical trajectory, i.e., for $\delta(T=0)=0$  scaling
yields
\begin{equation}
\xi\sim|\delta(T)|^{-\nu}\overset{\delta(0)=0}{\longrightarrow}\xi\sim
T^{-\nu/\psi}=T^{-1/2}
\end{equation}
so that at the QBP the correlation length diverges with decreasing
temperature as $1/\sqrt{T}$. The scaling of other physical
quantities can be predicted similarly, particularly, the specific
heat along this line has a non-Fermi liquid, $C_{P}(T)
\propto\sqrt{T}$, behavior~\cite{Mucio4}~(see Fig.~\ref{bicri}).

\section{Conclusions}

In this paper we have considered a generalized Ginzburg-Landau
functional including both superconducting and magnetic order
parameters to study the mutual influence of these instabilities in
three dimensional heavy fermion metals at zero temperature. The
quantum corrections to this classical functional were obtained
using the effective potential method up to one loop order. This
method had to be generalized to take into account the dissipative
nature of the propagator of the paramagnons in the metallic
paramagnetic phase close to the antiferromagnetic instability. The
superconductor propagator we used corresponds to the simplest
generalization of the Ginzburg-Landau classical action to take into
account dynamic or quantum effects. The calculations however can be
carried out with
different propagators, for example, that of Ramazashvili and
Coleman~\cite{Ramazashvili}. The results will be
presented elsewhere. We point out that the exact form of
the superconductor propagator is irrelevant for the qualitative
results of sections VI-A and VII.

Different phase diagrams have been considered according to the
distance between the quantum critical points in the classical
potential. For non-coincident QCPs the second order nature of the
quantum transitions obtained at the classical level is modified by
the quantum corrections, changing from continuous to weak first
order transitions. Hence, the spin fluctuations can change the
nature of the superconducting transition in the same way that the
coupling to a magnetic field does~\cite{Halperin,Mucio2}. For this
fluctuation induced first order transition we can calculate the
equivalent of the {\em latent heat} to transitions at $T=0$ and
the spinodal points. Both indicate that we are dealing with a weak
first order transition as they are directly related to the small
coupling constant $u$.

We also have found that a metallic antiferromagnetic QCP is
strongly affected by the presence of superconducting fluctuations.
These lead to the appearance of a small moment phase which goes
through a weak first order transition to a large moment one as the
system moves away from the superconductor instability.

Finally, in the case of coincident quantum critical points the
transition remains second order and the finite temperature
behavior can be extracted from the scaling properties of the
quantum bicritical point.

The effects of the competition between the AF and SC instabilities are highly
non-trivial since a positive coupling which classically tends to avoid the
coexistence of these states leads when quantum corrections are considered to
symmetry breaking in the normal state and
increases the region of the ordered phases (SC and AF) in the phase diagram.

In recent years increasing experimental evidence is being gathered
that points to existence of intrinsic inhomogeneities close to the
$T=0$ antiferromagnetic instability in heavy fermions~\cite{Flouquet}.
Coexistence of
small ordered moments with paramagnetic regions, coexistence
between superconductivity and paramagnetism have been observed.
The present approach provides an appealing model for this type of
phenomena. They arise in our theory as a direct consequence of the
competition between different order parameters which give rise,
through quantum corrections, to weak first order transitions with
the associated physics of metastability, spinodal points and phase
coexistence.

\begin{acknowledgments}
The authors would like to thank the Brazilian Agencies, FAPERJ and
CNPq for financial support.
\end{acknowledgments}

\end{document}